# Dual-Frequency Sub-Doppler Spectroscopy: Extended Theoretical Model and Microcell-Based Experiments

Denis V. Brazhnikov[1,2]*, Michael Petersen[3], Grégoire Coget[3], Nicolas Passilly[3], Vincent Maurice[3,4], Christophe Gorecki[3] and Rodolphe Boudot[3,4]**

[1]*Institute of Laser Physics SB RAS, Novosibirsk, Russia*
[2]*Novosibirsk State University, Novosibirsk, Russia*
[3]*FEMTO-ST, CNRS, UBFC, ENSMM, Besançon, France*
[4]*now at NIST, Time-Frequency Division, Boulder, Colorado, USA*
*brazhnikov@laser.nsc.ru, **rodolphe.boudot@femto-st.fr

Doppler-free spectroscopy using two counter-propagating dual-frequency laser beams in alkali vapor cells has been demonstrated recently, providing the detection of high-contrast sign-reversed natural-linewidth sub-Doppler resonances. However, to date, only a qualitative theory based on a simplified Λ-scheme model has been reported to explain underlying physics of this phenomenon. In this work, we develop a general and extended theoretical model of dual-frequency sub-Doppler spectroscopy (DF SDS) for Cs $D_1$ line. The latter considers the real atomic energy structure, main relaxation processes and various nonlinear effects including optical pumping, optical transition saturation, Zeeman and hyperfine coherent population trapping (CPT) states. This model allows to describe quantitatively the respective contributions of involved physical processes and consequently to estimate main properties (height and linewidth) of detected sub-Doppler resonances. Experimental results performed with a Cs vapor micro-fabricated cell are reported and explained by theoretical predictions. Spatial oscillations of the sub-Doppler resonance amplitude with translation of the reflection mirror are highlighted. Reported results show that DF SDS could be a promising approach for the development of a fully-miniaturized and high-performance optical frequency reference, with applications in various compact quantum devices.

## I. INTRODUCTION

Doppler-free spectroscopy in atoms and molecules [1,2] provides an exquisite tool to perform numerous high-precision measurements. This elegant technique has been demonstrated using several approaches including saturated absorption and dispersion spectroscopy [3-6], polarization spectroscopy [7], dichroic atomic-vapor laser lock (DAVLL) [8], selective reflection [9-11] or two-photon spectroscopy [12-14].

Saturated absorption spectroscopy has revealed to be a very helpful technique to explore fundamental aspects including the accurate measurement of molecular spectra [3,4,15,16], the observation of quantum fields and special relativistic effects [17-21] or the study of atomic and molecular kinetic processes [22-24].

Simultaneously, due to its relative simplicity and reliability, Doppler-free spectroscopy has been widely used in many physical experiments for laser frequency stabilization. Fractional frequency stabilities in the $10^{-13}$–$10^{-11}$ range at 1 s integration time have been demonstrated with cell-stabilized lasers using saturated absorption spectroscopy [25-27]. Based on similar physics and technologies, lasers frequency-stabilized to molecular lines have demonstrated remarkable performances with instability levels in the $10^{-14}$ and $10^{-15}$ range at 1 s and $10^4$ s respectively [28-29]. These results have induced great success for the production of compact optical frequency standards, including their recent deployment in space missions [30], optical communications [31], multi-wavelength laser interferometry [32], portable length standards [33] and compact fiber frequency synthesizers [34].

The usual light-field configuration for saturated-absorption spectroscopy (SAS) is based on two counter-propagating waves of same optical frequency $\omega$, traveling in a vapor cell filled with atoms or molecules. The natural-linewidth Doppler-free resonance can be detected using a photodiode right after the cell as a transparency peak in the bottom of a Doppler-broadened absorption profile when $\omega$ is scanned around the atom optical transition frequency $\omega_0$ or around a middle point $(\omega_{01}+\omega_{02})/2$ between two transition frequencies (so-called crossover resonances).

In a recent study [35], dual-frequency sub-Doppler spectroscopy (DF SDS) has been demonstrated to allow the detection of sign-reversed enhanced-absorption Doppler-free resonances. Compared to the usual single-frequency sub-Doppler spectroscopy (SF SDS), this approach has allowed to improve significantly the frequency stability of a diode laser [35], contributing to improve the performance of Cs cell atomic clocks [36,37].

A theoretical analysis of the DF SDS technique has been reported in [38]. This study has demonstrated that the detection of the high-contrast sign-reversed sub-Doppler resonances results from several complex physical phenomena including coherent population trapping (CPT) states of Zeeman sub-levels inside a single hyperfine (hf) state and between two hf-states and velocity-selective optical pumping effects. However, this theoretical analysis was based on a simplified three-level $\Lambda$-scheme model, only considering independently a few non-linear optical effects and then restricting to a limited qualitative understanding of the phenomenon.

In the present article, a general and extended theoretical model, considering the real energy structure of the atom with separate Zeeman sub-levels and the simultaneous contribution of various physical mechanisms and relaxation processes, is reported. The latter, at the opposite of the simplified model proposed in [38], allows to describe quantitatively the contribution of respectively-involved physical processes and consequently to simulate the properties (linewidth, height) of sub-Doppler resonances, including their dependence to main experimental parameters (laser intensity, position of the reflection mirror). An important result of these calculations is to predict that the use of short-length cells is a preferable configuration for DF SDS since the sub-Doppler resonance height can be maximized with proper position of the retro-reflection mirror. These spatial oscillations are attributed to interference effects between hyperfine CPT states induced by respective counter-propagating waves.

In order to evaluate the validity of our extended model, experimental tests are performed using a Cs vapor micro-fabricated cell. The impact of key experimental parameters on the sub-Doppler resonance properties is experimentally studied and found to be well-explained by the theoretical model.

The dependence of the resonance height on the mirror position, predicted by theory and intrinsically linked to the use of a short-length cell, is clearly demonstrated. We believe that high-contrast resonances detected using DF SDS could be of interest for the development of fully-miniaturized high-performance optical frequency references, with applications in various atomic devices and instruments.

## II. THEORY

### A. Problem statement

We consider an evacuated (no buffer gas) vapor cell placed in the field of two laser beams propagating in opposite directions along the quantization axis $z$ (see sketch in Fig. 1). Each of the beams in turn consists of two monochromatic plane waves:

$$\mathbf{E}(z,t) = \left[ E_1 \xi_1 e^{ik_1 z} + E_3 \xi_3 e^{-i(k_1 z + \phi_1)} \right] e^{-i\omega_1 t}$$
$$+ \left[ E_2 \xi_2 e^{ik_2 z} + E_4 \xi_4 e^{-i(k_2 z + \phi_2)} \right] e^{-i\omega_2 t} + \text{c.c.} \quad (1)$$

with $E_i$ the real amplitudes of the waves, $\phi_1$ and $\phi_2$ the phases of two backward waves, $\xi_i$ the unit complex vectors of the light wave polarizations, $k_{1,2} = \omega_{1,2}/c$ the wave numbers for the light waves with angular optical frequencies $\omega_{1,2}$, and "c.c." means the complex conjugate terms. It is assumed that the light waves have linear polarizations so that the components $\mathbf{E}_1$ and $\mathbf{E}_2$ are polarized along the $x$ axis, while polarizations of the other two waves ($\mathbf{E}_3$ and $\mathbf{E}_4$) are oriented at an angle $\alpha$ with respect to the $x$ axis. Therefore, in the spherical basis, we can write [39]:

$$\xi_{1,2} = (\mathbf{e}_{-1} - \mathbf{e}_{+1})/\sqrt{2}, \quad (2)$$

$$\xi_{3,4} = \left( e^{i\alpha} \mathbf{e}_{-1} - e^{-i\alpha} \mathbf{e}_{+1} \right)/\sqrt{2}, \quad (3)$$

where complex vectors $\mathbf{e}_{\pm 1}$ are spherical basis vectors responsible for $\sigma_+$ and $\sigma_-$ optical dipole transitions in the atom.

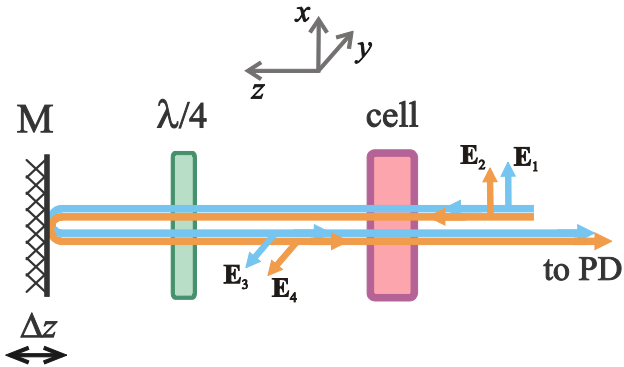

FIG. 1. Sketch of the proposed optical configuration: M: movable mirror, $\lambda/4$: quarter-wave plate, PD: photodiode.

Polarized light waves induce electric dipole transitions in alkali atoms, as shown in Fig. 2. Note that, for simplicity, Fig. 2 does not reflect the degeneracy of hyperfine levels over magnetic Zeeman sub-levels with quantum numbers $m_a = -F_a, -F_a+1,...$ with $F_a$ is the total angular momentum of "$a$" hyperfine level ($a$ = 1, 2, 3). This interaction between atoms and the light field leads to various nonlinear optical effects, such as optical pumping, optical transition saturation, coherences between magnetic sub-levels, and spontaneous anisotropy transfer from the excited state to the ground state. In the configuration considered here, a moving atom experiences a four-frequency light field, which induces multiple spatial harmonics of the atom's polarization. The finite size of the light beams leads also to time-of-flight relaxation. Our model includes all these effects in order to adequately reproduce the experimental observations.

The theoretical analysis is based on the standard density matrix formalism for a single atom, moving in gas. Interactions between atoms at low pressure gas do not affect seriously the signal and are omitted. The kinetic equation for the atom density matrix $\hat{\rho}$ has the Lindblad form (e.g., see [5,40]):

$$\left( \frac{\partial}{\partial t} + \upsilon \nabla_z \right) \hat{\rho} = -\frac{i}{\hbar} \left[ \left( \hat{V} + \hat{H}_0 \right), \hat{\rho} \right] + \hat{\Re}[\hat{\rho}]. \quad (4)$$

Here $\nabla_z = \partial/\partial z$, $\upsilon$ is the projection of the atom velocity on the $z$ axis.

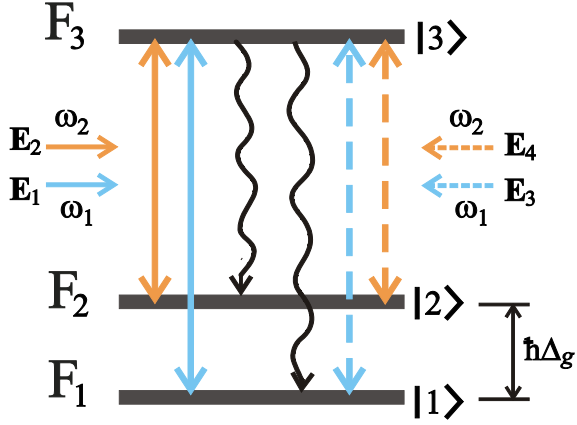

FIG. 2. Relevant energy levels of the $D_1$ line of an alkali atom. Degeneracy of energy levels $F$ over magnetic sub-levels $m$ is not shown. Solid arrows denote optical light-induced transitions by the waves propagating along the $z$-axis, while dashed arrows stand for counter-propagating waves. Wavy arrows are for spontaneous relaxation. $\hbar\Delta_g$ is the energy hyperfine splitting of the atom ground-state. The case depicted here corresponds to the atoms at rest under the null Raman ($\delta_R$) as well as one-photon ($\delta$) optical frequency detunings.

The operator $\hat{V} = \hat{V}_E + \hat{V}_B$ in (4) describes the interaction between the light waves ($E$) and the static magnetic field ($B$) in the electric-dipole approximation. $\hat{H}_0$ is the part of the total Hamiltonian for a free atom. The linear functional $\hat{\Re}[\hat{\rho}]$ in (4) is responsible for various relaxation processes in the atom, including the spontaneous relaxation described by the $\gamma$ constant and the transit-time relaxation taken into account by the $\Gamma$ constant ($\Gamma \approx \bar{v}/\tau$ with $\bar{v}$ the mean thermal velocity of an atom in gas, $\tau$ the mean time of atom's flight through the light field). By introducing the latter constant, we can omit derivatives over the transverse coordinates $\nabla_x$ and $\nabla_y$ in (4). Strictly speaking, this approach corresponds to light beams having step-like intensity cross sections. However, this approach remains acceptable and is widely used in theory with Gaussian-like profiles, considering a proper choice of the $\Gamma$ value. The light-induced recoil effect is not relevant for our study and is neglected.

All explicit expressions of the operators included in (4) are reported in the Appendix section.

The density matrix can be expanded into series of nine matrix blocks:

$$\hat{\rho} = \sum_F \hat{\rho}_{ab}(z,t)|F_a\rangle\langle F_b| \quad (a, b = 1, 2, 3), \quad (5)$$

where angular brackets stand for the Dirac bra- and ket-vectors. The diagonal blocks $\hat{\rho}_{aa}$ in (5) are responsible for magnetic sub-level populations of a single $|a\rangle$ level and coherent superpositions of these sub-levels (Zeeman coherences). $\hat{\rho}_{13}$, $\hat{\rho}_{23}$ and Hermitian conjugate matrices $\hat{\rho}_{31} = \hat{\rho}_{13}^\dagger$, $\hat{\rho}_{32} = \hat{\rho}_{23}^\dagger$ are known as optical coherences since they oscillate in time at optical frequencies. Finally, $\hat{\rho}_{12}$ and $\hat{\rho}_{21} = \hat{\rho}_{12}^\dagger$ are named hyperfine (hf) coherences since they oscillate in time at frequencies close to $\Delta_g$ with $\hbar\Delta_g$ the hyperfine energy splitting of the atom's ground state (see Fig. 2).

The proposed light-field configuration leads to an atom's polarization having a complex dependence on $z$-coordinate. In other words, the matrix blocks $\hat{\rho}_{ab}(z,t)$ can be expanded into series of various spatial harmonics. Following the work presented in [38], we only consider the lowest spatial harmonics. In the rotating wave approximation (RWA), we have the series:

$$\hat{\rho}_{aa}(z) \approx \hat{\rho}_{aa}^{(0)} + \hat{\rho}_{aa}^{(+)} e^{2ik_{12}z} + \hat{\rho}_{aa}^{(-)} e^{-2ik_{12}z}, \quad (6)$$

$$\hat{\rho}_{12}(z,t) \approx e^{i\delta_{12}t}\left(\hat{\rho}_{12}^{(+)} e^{ik_{12}z} + \hat{\rho}_{12}^{(-)} e^{-ik_{12}z}\right), \quad (7)$$

$$\hat{\rho}_{21}(z,t) \approx e^{-i\delta_{12}t}\left(\hat{\rho}_{21}^{(+)} e^{ik_{12}z} + \hat{\rho}_{21}^{(-)} e^{-ik_{12}z}\right), \quad (8)$$

where $k_{12}=k_1-k_2$ and $\delta_{12}=\omega_1-\omega_2$. Taking into account that (7) and (8) must be Hermitian conjugate to each other, we come to obvious properties: $\hat{\rho}_{21}^{(+)\dagger} = \hat{\rho}_{12}^{(-)}$ and $\hat{\rho}_{21}^{(-)\dagger} = \hat{\rho}_{12}^{(+)}$. For shortness, series expansions for the optical coherences are reported into the Appendix section

(see also [5] on the spatial-harmonics series expansion for elements of the density matrix). The series expansions (6)–(8) and (A13), (A14) in Appendix allow to consider all optical effects relevant for the present study. Note that, strictly speaking, these approximate expressions are valid only for low laser intensity levels.

We note that our model implies some approximations and simplifications. In particular, many of fast spatial oscillations that can be induced by simultaneous action of counter-propagating waves as, for instance, $e^{\pm ik_1 z \pm ik_2 z}$, $e^{\pm 2ik_1 z}$, $e^{\pm 2ik_2 z}$ and others, are dropped from our consideration. Such fast spatial oscillations come from simultaneous action of counter-propagating waves on atomic polarization and may have noticeable influence on the spectroscopic signals, even for moderate light-field intensities of several mW/cm$^2$ (for dipole transitions in alkali atoms). The influence of high-order spatial harmonics on sub-Doppler resonances was studied in several works [1,41], including the possibility to destroy some nonlinear optical effects well-observed under weak fields [42]. In our experimental study, moderate to high intensities (up to 500 mW/cm$^2$) are used. However, the contribution of higher-order spatial harmonics has not been taken into account in the present study for two reasons. First, considering them would have complicated dramatically our analysis and especially numerical calculations. Second, experimental intensities of both counter-propagating light beams cannot be equal to each other due to losses of optical elements of the setup and light-field absorption in the cell. The latter reason is more especially confirmed in the DF regime where absorption is enhanced. Thus, the influence of higher-order spatial harmonics is significantly suppressed and we can keep only slow oscillating terms such as $e^{\pm ik_1 z \mp ik_2 z}$.

From (6), it is seen that the temporal evolution of sub-level populations is not considered. This is explained by the fact that all the transient processes are assumed to be completed. The so-called steady-state approximation of light-atom interaction is then used. Another approximation is that we neglect the light-field interaction with the second excited-state hyperfine level $F_4$ (not shown in Fig. 2) of the D$_1$ line. This simplification can be validated since the hyperfine energy splitting in the excited state for Cs atom is large enough ($\Delta_e \approx 2\pi \times 1.17$ GHz), compared to the typical Doppler linewidth (FWHM$_{Doppler} \approx 2\pi \times 370$ MHz). We assume here that $F_3 = F_2 = F_1 + 1$, with $F_1 = 3$, $F_2 = F_3 = 4$ for the Cs D$_1$ line.

The light-field intensity change due to absorption in the cell can be written formally as the Beer-Lambert law:

$$I_t(z_c) = I_{0t} e^{-\text{OD}} \quad (9)$$

where $I_{0t}$ is the total intensity before the cell and OD is the optical density of the medium such that:

$$\text{OD} = -\int_{z_c}^{z_c + L} \chi(z) dz \, , \quad (10)$$

with $\chi$ the absorption coefficient for the total light-field in the cell and $z_c$ the position of the cell face window along the $z$-axis. Note that we do not need to integrate twice over $L$ since we consider the absorption coefficient for the total light field in the cell.

The absorption coefficient depends on many parameters such as the optical frequency detuning $\delta$ and the two-photon (Raman) detuning $\delta_R$, the coordinate $z$ within the cell, the cell position $z_c$, the polarization angle $\alpha$ and intensities of all light waves $I_{1,2,3,4}(z)$, being also functions of the $z$ coordinate. Besides, $\chi$ depends on the phases $\phi_1$ and $\phi_2$ of the counter-propagating waves. It can be easily shown that $\chi$ depends only on their difference $\phi_{12} = \phi_1 - \phi_2$.

Instead of considering the real dependence $\chi=f(z)$ and solving the complicated Maxwell-Bloch system of equations, we use instead of (10) a proper approximate expression explained by the following assumptions. First, the alkali vapor is considered to be optically thin, i.e. OD<<1. In the experiments, the optical density OD can be controlled by adjusting the cell temperature. Secondly, the coefficient $\chi$ is determined by the total population of the atom's excited state $\text{Tr}[\hat{\rho}_{33}(z)]$ averaged over the Maxwellian velocity distribution. As follows from (6), the population undergoes spatial variations due to nonlinear interference effects. Thus, $\chi$ should reflect the same oscillations. However, the cell length $L$ is assumed to be much smaller than the period of these oscillations, i.e.

$$T_z = \pi/k_{12} = \pi c/\Delta_g \approx 16.3 \text{ mm} \gg L \approx 1.4 \text{ mm} \quad (11)$$

Consequently, assuming a small optical density, the light intensity recorded by the photodetector can be written as:

$$I_t(z_c) \approx \eta I_{0t} e^{-\chi(z_c)L} \approx \eta I_{0t}(1-\chi(z_c)L), \quad (12)$$

where $\eta$ stands to consider possible intensity losses on optical elements of the setup. The absorption coefficient can be expanded into two parts:

$$\chi \propto W_e = \frac{1}{\sqrt{\pi}\,\upsilon_0} \int_{-\infty}^{\infty} \text{Tr}[\hat{\rho}_{33}(z_c,\upsilon)] e^{-\upsilon^2/\upsilon_0^2}\, d\upsilon$$

$$= W_0(\delta,\alpha,I_{1-4}) + W_z(\delta,\alpha,\phi_{12},I_{1-4},z_c)\,, \quad (13)$$

with:

$$W_0 = \left\langle \text{Tr}\left[\hat{\rho}_{33}^{(0)}\right] \right\rangle_\upsilon, \quad (14)$$

and

$$W_z(z) = 2\,\text{sinc}(Lk_{12})$$
$$\times \left\langle \text{Re}\left\{ \text{Tr}\left[\hat{\rho}_{33}^{(+)}\right] e^{ik_{12}(2z_c+L)} \right\} \right\rangle_\upsilon. \quad (15)$$

Here "sinc" is the un-normalized sinc function. The velocity $\upsilon_0 = \sqrt{2k_B T/m_a}$ in (13) is the most probable atom thermal velocity with $k_B = 1.38\times 10^{-23}$ J/K the Boltzmann constant and $m_a$ the atom's mass. Brackets $\langle...\rangle_\upsilon$ stand for averaging over the Maxwellian velocity distribution.

In SDS, the light field transmitted through the vapor cell is monitored as a function of the optical frequency. In our case, this is equivalent to scanning $I_t$ over $\delta = \omega_0 - (\omega_{31}+\omega_{32})/2$, corresponding to the optical frequency detuning of the laser carrier $\omega_0 = (\omega_1+\omega_2)/2$ from the middle frequency of both optical transitions $|1\rangle \to |3\rangle$ and $|2\rangle \to |3\rangle$ (see Fig. 2). As long as $W_e$ determines all the nonlinear optical effects observed in $I_t$, we will analyze $W_0(\delta)$ and $W_z(\delta)$ separately for different physical conditions. Let us just note that dividing expression (13) into two parts has a real physical meaning: the second term $W_z$ is only responsible for the effects caused by hf coherences, while the first term $W_0$ reflects all the other effects including optical pumping, optical transition saturation and CPT within a single level $|2\rangle$. The latter CPT effect is called Zeeman-CPT since it embraces Zeeman sublevels of a single hyperfine $F$ level. Note additionally that as long as we consider the transition $F_1 \to F_3=F_1+1$ which is not a transition of the "dark" type [43], the CPT does not occur within the $|1\rangle$ level. This simplifies a little our analysis.

### B. Spatial oscillations of the sub-Doppler resonance height: Qualitative treatment

Spatial oscillations described in (6) play an important role in our study. The origin of these oscillations can be easily understood on the basis of a simple $\Lambda$-scheme, considered briefly in the following. We assume intensities of co-propagating light waves to be equal, i.e. $I_1=I_2$ and $I_3=I_4$. In a first step, we consider a non-degenerate ground state, so that energies of levels $|1\rangle$ and $|2\rangle$ are different since they belong to different ground-state hyperfine components as shown in Fig. 2.

Owing to the linear Doppler effect, the two light waves $\mathbf{E}_1$ and $\mathbf{E}_2$ interact with a resonant velocity group of atoms having a velocity $\upsilon \approx \delta/k$ (in this expression, the difference between $k_1$ and $k_2$ is not relevant and can be neglected). If the off-resonance condition $\delta \gg \Delta_{\text{res}}$ is satisfied (with $\Delta_{\text{res}}$ the FWHM of the sub-Doppler resonance), the resonant group of atoms does not "experience" considerable influence of the other two light waves $\mathbf{E}_3$ and $\mathbf{E}_4$. If the two-photon (Raman) detuning equals zero, i.e. $\delta_R=\omega_1-\omega_2-\Delta_g=0$, and if the light field components $\mathbf{E}_1$, $\mathbf{E}_2$ are mutually coherent, then atoms can be pumped into a "dark" (non-coupled) state [44,38]:

$$|\text{NC}_1\rangle = \frac{1}{\sqrt{2}}\left(|1\rangle - e^{ik_{12}z}|2\rangle\right) . \quad (16)$$

The spatial phase $e^{ik_{12}z}$ comes from the different propagation phases of both waves $\mathbf{E}_1 \propto e^{ik_1 z}$ and $\mathbf{E}_2 \propto e^{ik_2 z}$. This dark state leads to a low level of light-field absorption in the medium [44, 45].

At the same time, $\mathbf{E}_3$ and $\mathbf{E}_4$ waves pump the atoms with $\upsilon' \approx -\delta/k$ into another non-coupled state:

$$|\text{NC}_2\rangle = \frac{1}{\sqrt{2}}\left(|1\rangle - e^{-i(k_{12}z+\phi_{12}+2\alpha)}|2\rangle\right) , \quad (17)$$

Similarly, light waves $\mathbf{E}_3$ and $\mathbf{E}_4$ experience small absorption in the cell due to the CPT phenomenon for the resonant velocity group of atoms $\upsilon' \neq \upsilon$.

Let us now consider the in-resonance regime $\delta \leq \Delta_{\text{res}}$. In this case, all the light waves $\mathbf{E}_{1,2,3,4}$ interact with the same atoms ($\upsilon'=\upsilon$) leading to some kind of "competition" between (16) and (17) states. The result of this competition depends on several factors and leads to the observation of increased or decreased level of the light-field absorption at the resonance profile center. The decreased vapor absorption (increased vapor cell transmittance), i.e. the regular saturated-absorption transparency peak, takes place when both non-coupled states are "parallel":

$$\langle \text{NC}_1 | \text{NC}_2 \rangle = 1 . \quad (18)$$

As seen from (16) and (17), this may happen when:

$$k_{12}z + \alpha + \frac{\phi_{12}}{2} = \pi n . \quad (19)$$

The "parallelism" of $|\text{NC}\rangle$ states means that both states are dark states for the total light field in the cell. At the opposite, if both non-coupled states are orthogonal, i.e.

$$\langle \text{NC}_1 | \text{NC}_2 \rangle = 0 , \quad (20)$$

the reduction of absorption due to CPT does not occur and the medium intensively scatters photons from the resonant light field, leading to significantly increased absorption (or reduced transmission) of the cell. This happens when:

$$k_{12}z + \alpha + \frac{\phi_{12}}{2} = \frac{\pi}{2}(1+2n) . \quad (21)$$

In particular, if the polarization configuration of the waves is fixed ($\alpha=\pi/2$) as well as the relative phase $\phi_{12}$, the observation of increased or decreased vapor cell absorption will then depend on the position of the cell along the $z$ axis.

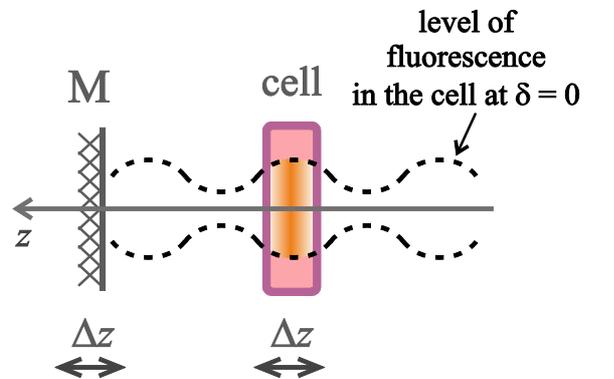

FIG. 3. Sketch to demonstrate spatial oscillations of light scattering during mirror or vapor cell position change.

To illustrate this, a sketch is drawn on Fig. 3 where nodes and antinodes of the light-induced fluorescence depend on the cell position. To check this effect in experiments, the cell position can be fixed at $z=z_c$, while the position of the mirror "M" is changed. This approach is equivalent to change the phase $\phi_{12}$.

Spatial oscillations discussed here are attributed to hyperfine CPT states in contrast to Zeeman-CPT states which do not depend on the phase $\phi_{12}$. Indeed, an optical dipole transition between two degenerate energy levels $F_a=F \rightarrow F_3=F$ ($a = 1$ or $2$) can be also reduced to a regular $\Lambda$-scheme to perform a qualitative theoretical analysis.

In contrast to the previous case, the two ground sub-levels $|1\rangle$ and $|2\rangle$ have equal energies, belonging to a common hyperfine level $F$ ($F=4$ for Cs atom). These levels are excited by light waves having wave vectors with equal absolute values. This circumstance leads to the following non-coupled states:

$$|NC_1\rangle = \frac{1}{\sqrt{2}}(|1\rangle - |2\rangle), \quad (22)$$

$$|NC_2\rangle = \frac{1}{\sqrt{2}}(|1\rangle - e^{-2i\alpha}|2\rangle), \quad (23)$$

with $\alpha$ is angle between linear polarizations of counter-propagating waves. Similarly to the qualitative analysis provided before, both states (22) and (23) can be parallel or orthogonal depending on the angle $\alpha$. In other words, this angle controls the strength of the light field absorption when both counter-propagating waves act on the same atoms, leading to the possible observation of a sub-Doppler dip or peak at the center of the resonance profile. However, the sign of the resonance does not depend now on the position of the vapor cell or the mirror. The corresponding reason for the sub-Doppler resonance's sign change can be attributed to the Zeeman-CPT effect, because now $|1\rangle$ and $|2\rangle$ model two magnetic sub-levels of a single hyperfine level $F$ in contrast to the hf-CPT effect.

We have shown that Zeeman-CPT as well as hf-CPT effects can provide the observation of the sub-Doppler resonance with enhanced absorption. An obvious prospect is then to predict how to make these two nonlinear effects work and add together. The simple $\Lambda$-scheme considered in this subsection and in previous studies [35, 38] cannot help to solve this problem. Thus, considering the real structure of atomic energy levels is a natural and rigorous approach to take into account the contribution of both Zeeman-CPT and hf-CPT effects. This approach will also reflect the influence of other nonlinear effects such as for example spontaneous anisotropy transfer from excited to ground states. The precise consideration is based on the density matrix equations (see Section II-A and Appendix). The results of numerical calculations for the real structure of atomic energy levels and their brief discussions will be provided in the next section.

We should note that the analysis and effects presented in this subsection have many in common with those discussed in [46-48] for the hf-CPT effect and in [49,50] for the Zeeman-CPT effect. However, these previous works aimed to study sub-natural electromagnetically induced transparency and absorption (EIT/EIA) resonances observed by scanning the Raman detuning. In the present study, we focus on sub-Doppler natural-linewidth resonances.

It is often convenient to work with normalized quantities. We define $A_{norm}$ as the sub-Doppler resonance height $A$ normalized by the wide Doppler profile height $A_D$. On the basis of eq. (13), $A_{norm}$ is defined as:

$$A_{\text{norm}}(z_c, \phi_{12}) \approx (A_0 + A_z)/A_D, \quad (24)$$

where $A_0$ is the height of a sub-Doppler resonance in function $W_0(\delta)$, while $A_z$ is the height of the resonance in $W_z(\delta)$, so that the total height is just a sum $A=A_0+A_z$ (see Fig. 4, solid orange curve). The Doppler profile height $A_D$ in (24) is contributed only by a broad background profile in function $W_0(\delta)$, because $W_z(\delta)$ contains only a homogeneously broadened sub-Doppler resonance, as demonstrated in the next subsection.

Here, we see that the proposed expansion of the excited state population in (13) into two main parts help to understand the main features of the resonance height spatial oscillations. The term $A_0/A_D$ in (24) is just the normalized height of the resonance in the function $W_0(\delta)$ and does not depend on the coordinate at all. This term describes the spatially averaged value of the height oscillations. The other term $A_z/A_D$ explains the dependence of $A_{\text{norm}}$ on the $z$-coordinate.

From eq. (15), it is clear that the height of the sub-Doppler resonance oscillates with $z_c$ and falls off with the cell length as $\text{sinc}(k_{12}L)$. This, optimization of the sub-Doppler resonance height requires both a short cell and a correct choice of $z_c$, as experimentally demonstrated in the following section.

### C. Analysis of the high-contrast effect

Let us analyze contributions $W_0$ and $W_z$ in (13), in order to reveal their physical meaning and their influence on the total light field absorption in the cell under different physical conditions. In our following calculations, we use typical parameters for such Doppler-free spectroscopy experiments [35,38], with the original specificity that a miniaturized cell is considered. We consider the real structure of Cs $D_1$ line with $\lambda$=894.6 nm, $\gamma$=2$\pi$×4.56 MHz, $F_1$=3, $F_2$=4 and $F_3$=4 (see Fig.2).

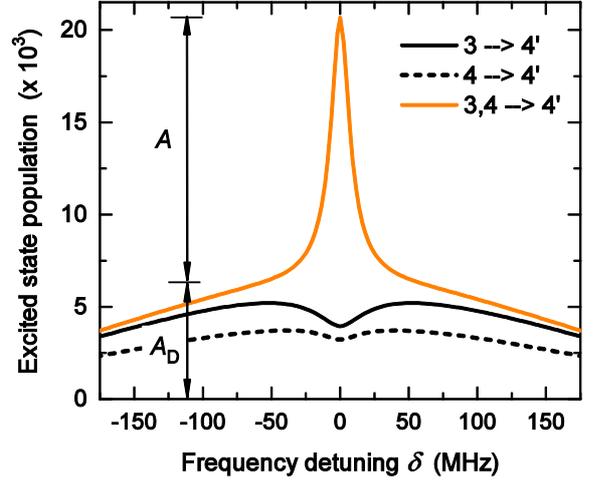

FIG. 4. Sub-Doppler resonances calculated for the single (solid and dashed curves at the bottom) and dual-frequency (solid red spike) regimes of light-field excitation. The total laser beam power at entrance of the cell is 50 μW. The magnetic field is switched off. Other parameters are written in the text.

All the levels are degenerate over magnetic (Zeeman) sublevels $m$=–$F$,–$F$+1,..., $F$. For all calculations, we take the time-of-flight relaxation rate $\Gamma = 0.02\gamma$, corresponding to a laser beam diameter of about 0.5 mm. The Doppler HWHM is $kv_0$=50$\gamma$. For figures 4 to 7, the magnetic field is null. The Raman frequency detuning under the DF regime is assumed to be zero ($\delta_R$=0). To obtain the resonance curves, the optical frequency detuning $\delta$ is scanned.

As shown previously, the absorption coefficient $\chi$ is proportional to the total excited-state population of the atom (13). Therefore, we focus on analyzing the population $W_e$ as a function of the frequency detuning $\delta$.

Figure 4 depicts numerically calculated resonances in both SF and DF regimes. When a single optical transition is excited (either $|1\rangle \rightarrow |3\rangle$ or $|2\rangle \rightarrow |3\rangle$, solid and dashed black curves), the regular saturated-absorption absorption dip is observed. The sign of the sub-Doppler resonance changes when the regime of excitation is switched to the DF regime, with $|3\rangle$ ($F_3$=4) being the common excited level. In addition to the change of the resonance sign, we observe that the latter becomes narrower and with a higher contrast.

*1. Influence of the light wave polarizations and phases*

We focus here on the analysis of both contributions $W_0$ and $W_z$ from (13), being defined as both parts of the absorption coefficient. Fig.5a reflects the influence of the polarization configuration of both contributions $W_0$ and $W_z$ on the total excited state population $W_e$. It is seen that both $W_0$ and $W_z$ depend in a relevant manner on the angle between linear polarizations of both counter-propagating laser beams.

We observe first that the wide Doppler background of $W_0$ does not depend on the angle at all. At the opposite, the background $W_z$ strongly depends on the angle, so that $W_z$ changes its sign. This behavior can be clearly understood from the qualitative analysis provided in Section II-B. In particular, when the optical detuning is large enough ($\delta \gg \Delta_{res}$), both counter-propagating dual-frequency laser beams interacting with different velocity groups of atoms do not "feel" the presence of each other. Moreover, atoms in both groups are pumped into the dark states. As far as these resonant groups of atoms have different velocities, there is no "competition" between the dark states $|NC_1\rangle$ and $|NC_2\rangle$, and both states survive. Note that there can be different types of dark states in each resonant group of atoms, but we do not specify them here. These dark states lead to a low level of light absorption at the Doppler "wings" of the absorption profile (see solid green and dash-dotted blue curves in Fig.5a).

At resonance ($\delta \leq \Delta_{res}$), both laser beams interact with the same atoms. The result of this interaction depends strongly on the polarization configuration, because the quantum state of the atom depends on the angle $\alpha$. For instance, the case $\alpha = \pi/2$ leads to a significant increase of $W_0$ and $W_z$. The increase of $W_z$ is due to the destruction of the hyperfine CPT effect considered in Section II-B. The vapor becomes then less transparent for light. The increase of the second contribution ($W_0$) is also caused by the destruction of the CPT state within the $|2\rangle$ level. In this case, two possible Zeeman dark states are orthogonal under orthogonally polarized configuration of counter-propagating beams.

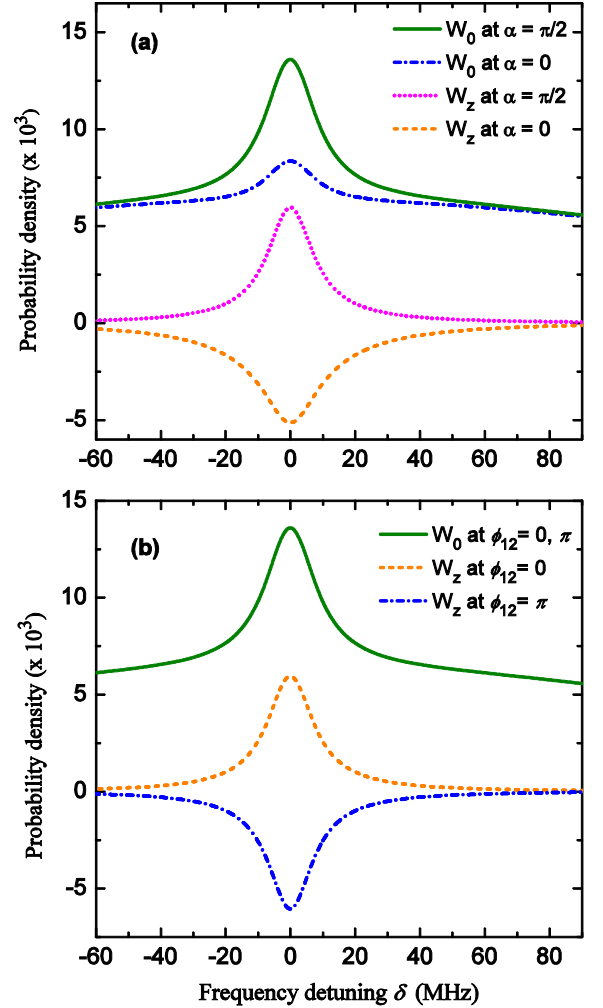

FIG. 5. Analysis of different contributions to the excited-state population. (a) Influence of linear polarizations orientation at mutual backward waves phase $\phi_{12}=0$. (b) Influence of the mutual phase $\phi_{12}$ for orthogonal linear polarizations of the counter-propagating waves ($\alpha=\pi/2$). The total laser beam power is 50 μW, the static magnetic field $B$ is switched off.

Using parallel linear polarizations, the dark states linked to magnetic sublevels of *different* hyperfine levels $|1\rangle$ and $|2\rangle$, i.e. hf-CPT states, do not compete with each other. At resonance ($\delta \leq \Delta_{res}$), atoms are pumped faster into hf-CPT states due to the simultaneous action of both beams. This process causes the creation of a dip in $W_z$ (see Fig.5a, dashed orange line). The same picture may be expected to occur with the $W_0$ term. Indeed, if the light beam polarizations are parallel ($\alpha=0$), two dark states $|NC_1\rangle$ and $|NC_2\rangle$, which could be created at $\delta \gg \Delta_{res}$ within the $|2\rangle$ level by independent light beams, should survive also at $\delta \approx \Delta_{res}$, because these states are parallel. This circumstance is expected to lead to the observation of a dip-like structure in the center of $W_0(\delta)$ as it occurs in the case of $W_z(\delta)$. However, we only observe (blue dash-dotted line in Fig.5a) a significant reduction of the height of the central resonance, while the sign is still positive (it is a peak). The decrease of the height can be explained by the absence of any competition between the different Zeeman dark states. The residual peak observation can be caused by optical pumping effects discussed further in the text.

Fig.5b shows the behavior of $W_0$ and $W_z$ for different values of the mutual phase $\phi_{12}$. It is seen that $W_0$ is immune to the change of $\phi_{12}$. This is explained by the fact that this phase influences the low-frequency (hf) coherences and corresponding nonlinear effects caused by these coherences. The phase is not relevant for any other effects which do not depend on hf-coherences such as optical pumping or Zeeman-CPT effects. The sign change observed in $W_z(\delta)$ is qualitatively explained in Section B.

Let us note that the dependence of $W_e$ on the polarization angle $\alpha$ and the relative phase $\phi_{12}$ can be treated in another way. Indeed, the angle $2\alpha$ is the relative phase between $\sigma_+$ and $\sigma_-$ transitions induced by the counter-propagating beams. If the in-resonance condition is satisfied ($\delta \leq \Delta_{res}$), the two counter-propagating light beams create various closed contours of atom-light interaction. The general theory of such interaction contours, provided in [51], revealed the strong sensitivity of the coherent phenomena to the light field phases ($\alpha$ and $\phi_{12}$ in our case).

### 2. Influence of the transition openness

Consider now the SF regime, when only the $|2\rangle \rightarrow |3\rangle$ transition is excited by both counter-propagating light waves (see Fig. 2). This approach will help to explain the absence of a sign-reversal effect in Fig.5a (for the solid green and dash-dotted blue curves) when the angle $\alpha$ changes from $\pi/2$ to 0. The other transition $|1\rangle \rightarrow |3\rangle$ with $F_1=3$ and $F_3=4$ is "bright" and cannot have any non-coupled (dark) state. In other words, it can be omitted when the Zeeman-CPT effect is considered. Otherwise, in the case where the single transition $|1\rangle \rightarrow |3\rangle$ is considered to be excited by counter-propagating waves under the SF regime, it will exhibit a regular saturated-absorption resonance at the center as a reduction of the light-wave absorption in the cell. Nevertheless, we will see later that the $|1\rangle \rightarrow |3\rangle$ transition plays the specific role of a population repumper. The latter cannot be excluded from our consideration and will be discussed further.

Fig.6 reports the evolution of $W_0$ in the SF regime when only the $|2\rangle \rightarrow |3\rangle$ transition is excited. The coefficient $\beta$ in the figure is the branching ratio. It characterizes the openness of the transition: $0 \leq \beta \leq 1$ and $\beta=1$ corresponds to a cyclic (closed) transition without any spontaneous decay to other non-resonant hyperfine levels (as $|1\rangle$). Fig. 6 shows that the sign-reversal effect should be observed when the transition is closed. In this condition, the Zeeman-CPT effect occurs and leads to the observation of a peak-like resonance.

At the opposite, if the transition is noticeably open (this is the case for $F_2$=4→$F_3$=4 in Cs, $\beta$=5/12), the Zeeman-CPT effect is significantly suppressed. In this case, both polarization configurations ($\alpha=\pi/2$ or 0) do not lead to any peak-like resonance observation. However, as shown on Fig.5a (solid green and dash-dotted blue curves), the Zeeman-CPT effect contributes to the peak observation in the DF regime when both optical transitions are excited. In other words, the Zeeman-CPT effect does not work when the SF regime is used and this effect is again in action when the DF regime is switched on. The latter happens owing to optical pumping of the atoms back to the $|2\rangle$ level by the second field resonant with the $|1\rangle \rightarrow |3\rangle$ transition. Thus, this second field plays the role of an optical repumper and increases the effective branching ration for $|2\rangle \rightarrow |3\rangle$ transition so that $\beta_{\text{eff}} > \beta = 5/12$. This explains why the resonance peak observed in $W_0(\delta)$ is so sensitive to the polarization angle change.

While the Zeeman-CPT effect contributes to the creation of the peak creation in $W_0$ in Fig.5a, another mechanism, the regular optical pumping effect, prevents the change of the resonance sign (compare solid green and dash-dotted blue curves in Fig.6). Indeed, it can be shown that the pumping rate of the field $\mathbf{E}_1+\mathbf{E}_3$ resonant to the $|1\rangle \rightarrow |3\rangle$ transition just slightly depends on the polarization angle $\alpha$ ($\pi/2$ or 0). Consequently, when both counter-propagating beams interact with the same atoms ($\delta \leq \Delta_{\text{res}}$), more atoms are pumped to the $F_2$=4 level and the absorption increases. As already mentioned, the level of light absorption on the $|2\rangle \rightarrow |3\rangle$ transition depends on $\alpha$ owing to the Zeeman-CPT effect. This is observed as a decrease or increase of the resonance height in Fig.5a in the $W_0(\delta)$ dependency.

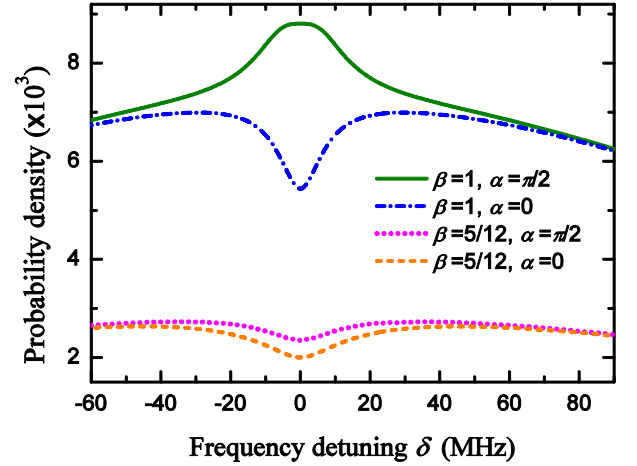

FIG. 6. Calculated contribution $W_0$ in the single-frequency regime, when only the transition $|2\rangle \rightarrow |3\rangle$ is excited. $\beta$ is the branching ratio for the transition. The total laser beam power is 50 μW. The static magnetic field is null.

*3. Influence of the imbalance between counter-propagating light wave intensities*

Figure 7 depicts the influence on $W_0$ and $W_z$ of the light wave intensities imbalance. Experimentally, this imbalance could come from the light field absorption in the cell after one pass and to losses caused by the imperfect optical elements.

The contribution of both hyperfine-CPT and Zeeman-CPT states to the absorption peak increase is optimized in the case where both counter-propagating laser beams have the same intensity. This condition is the best one to destroy the CPT state of the atom at the resonance center ($\delta \leq \Delta_{\text{res}}$) and then to increase the level of the light field absorption. It is well shown in Fig. 7 that the intensity imbalance affects the strength of the central peak. In all cases, the latter can be however well observed.

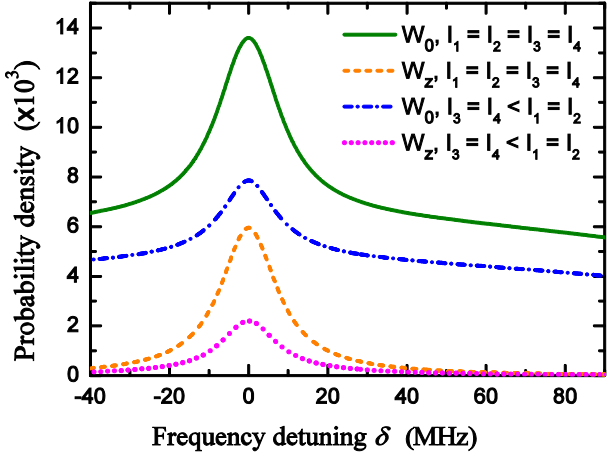

FIG. 7. Influence of a difference in light wave intensities on $W_0$ and $W_z$. Parameters: $\phi_{12}=0$, $\alpha=\pi/2$. The total laser beam power is 50 μW, $B=0$. Intensities $I_3$ and $I_4$ are assumed to be smaller than $I_1$ and $I_2$ by 30%.

*4. Influence of an ambient static magnetic field*

Figure 8 analyses the influence of a static magnetic field applied along the wave vectors ($\mathbf{B}\|z$) which is associated with the Larmor frequency $\Omega$. This frequency is different for different energy levels and is responsible for linear shifts of magnetic sublevels $m$ under the external magnetic field. Here we use a notation $\Omega\equiv\Omega_2=g_2\mu_B B/\hbar$ with $g_2$ the Landé g-factor of the $F_2$ level and $\mu_B=927.4\times10^{-26}$ J/T the Bohr magnetron. Other frequencies can be expressed via $\Omega$, because for the alkali atom $\Omega_1=-\Omega_2$ and $\Omega_3=(g_3/g_2)\Omega_2$.

As already discussed, the creation of the central absorption peak is due to the presence of the dark state at $\delta\gg\Delta_{res}$ and its absence at $\delta\lesssim\Delta_{res}$, when two orthogonal and linearly polarized counter-propagating light beams act on the same atoms. If a static magnetic field is applied, Zeeman sublevels of the $|2\rangle$ level are frequency shifted and the dark state is not created at all for any value of the one-photon detuning $\delta$. Obviously, the magnetic field destroys the dark state whatever the light wave polarizations. Consequently, we observe a high level of light-field absorption in both polarization configurations (blue dash-dotted curves in Fig.8a,b).

The presence of a static magnetic field leads to the creation of an absorption dip in the center of the absorption profile of $W_0$ term due to the regular saturated absorption effect. Concerning $W_z$ (see orange dashed and pink dotted curves in Fig.8b), the application of the magnetic field does not lead to the total destruction of the peak effect. The absorption peak effect in $W_z$ is still possible due to the fact that some Λ-schemes insensitive to the weak magnetic fields and some of the hyperfine CPT states can survive. These states may lead to manifestation of the hf-CPT effect and the absorption peak observation.

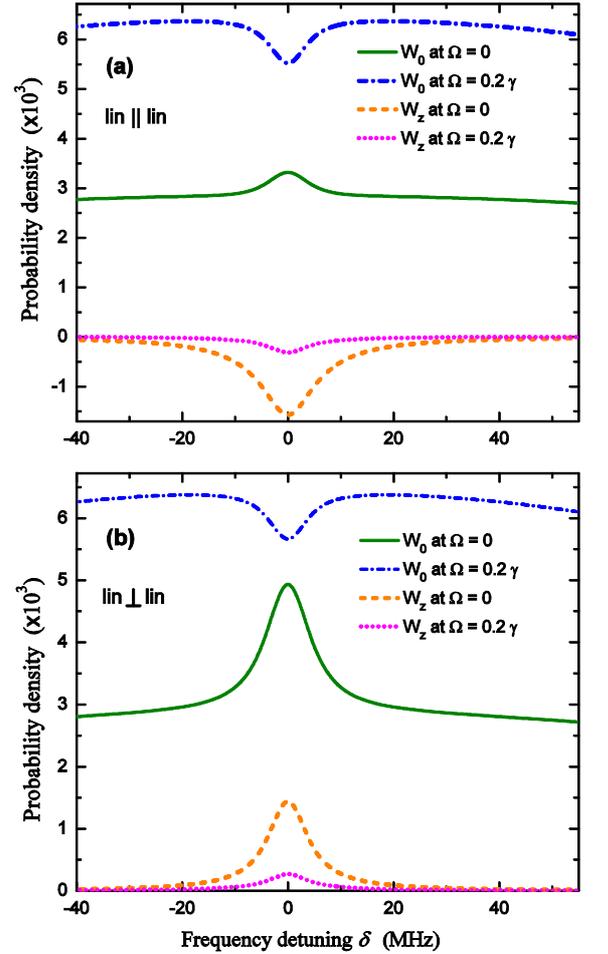

FIG. 8. Influence of the static magnetic field on $W_0$ and $W_z$. The field is applied along the light wave vectors ($z$ axis) at (a) parallel and (b) orthogonal linear polarizations of counter-propagating laser beams. Parameters are: $P=10$ μW and $\phi_{12}=0$. $\Omega$ is the Larmor frequency.

## III. MEASUREMENTS AND COMPARISONS WITH THEORY

### A. Setup

The heart of the experiment, shown on Fig. 9(a), is a Cs vapor micro-fabricated cell analog to the one described in [52-54]. It is made of two DRIE (deep reactive ion etching)-etched silicon cavities connected by thin channels and subsequently anodically-bonded to two borofloat glass wafers. The first cavity contains a Cs pill-dispenser laser-activated after the final cell bonding in order to generate Cs alkali vapor. Doppler-free spectroscopy takes place in the 2-mm diameter and 1.4-mm long cylindrical neighboring cavity. The cell does not contain any buffer gas. The cell is temperature-stabilized using a custom-made temperature controller and is placed inside a 6.5 cm long and 5 cm diameter cylindrical magnetic shield.

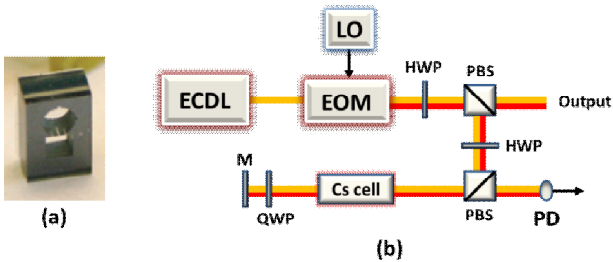

Fig. 9: (a): Photograph of a Cs vapor micro-fabricated cell. The square cavity is the dispenser cavity. The cylindrical cavity is the one where Doppler-free spectroscopy takes place. (b) Experimental setup for DF SDS measurements in the Cs microcell. ECDL: external-cavity diode laser, EOM: electro-optic modulator, LO: 4.596 GHz microwave synthesizer, HWP: half-wave plate, PBS: polarizing beam splitter, QWP: quarter-wave plate, M: mirror, PD: photodiode. The 4.596 GHz signal is applied to the EOM for the dual-frequency measurements. The first half-wave plate is used to control the laser power sent through the cell. The second half-wave plate helps to maximize the power sent to the arm with the micro-cell, ensuring the best possible linear polarization.

The vapor cell is used in the experimental setup shown in Fig. 9(b). The light source is a narrow-linewidth external cavity diode laser (ECDL, TopticaDL pro) tuned to the Cs $D_1$ line. The output beam is connected to a fibered intensity electro-optical modulator (EOM) via a polarization maintaining optical fiber. For dual-frequency Doppler-free spectroscopy tests, the EOM is modulated by a 4.596315885 GHz microwave signal with 22.8 dBm of power. This frequency is generated by a commercial microwave frequency synthesizer (Keysight E8257D) referenced to a local hydrogen maser.

For the standard single-frequency Doppler-free spectroscopy tests, the EOM is not modulated. In both cases, the EOM bias voltage is set to optimize the EOM output power. When applying microwave modulation, the optical carrier power is significantly reduced and two optical sidebands frequency-split by 9.192631770 GHz are generated. Before entering the cell, the light beam passes two half-wave plates and two polarizing beam splitters ensuring power control and linear polarization in the cell. The light beam is retro-reflected using a mirror through the cell for the detection of the sub-Doppler resonance. The mirror is placed onto a translation stage. A quarter-wave plate between the cell and the reflection mirror ensures the polarization of the reflected beam is orthogonal to the polarization of the incident light. At the output of the cell, the beam exits straight through the second beam splitter and is detected by an amplified photo-diode. The photodiode output signal is transferred to a digital oscilloscope connected to a computer. The width of the light beam is about 0.45 mm and is smaller than the cell diameter.

### B. Measurements

Figure 10 shows typical Doppler-free spectroscopy spectra detected in the micro-fabricated cell using the single-frequency (SF) or the dual-frequency (DF) regimes for a cell temperature of 60°C. In the SF regime, we observe the standard saturated absorption spectroscopy feature with increased transmission of the light through the vapor when the laser frequency is resonant with the atomic optical transitions. In the DF regime, as reported in [35] with cm-scale cells, a significant sign-reversal of the Doppler-free dip is observed and a narrow and high-contrast absorption spike can be observed with enough laser power.

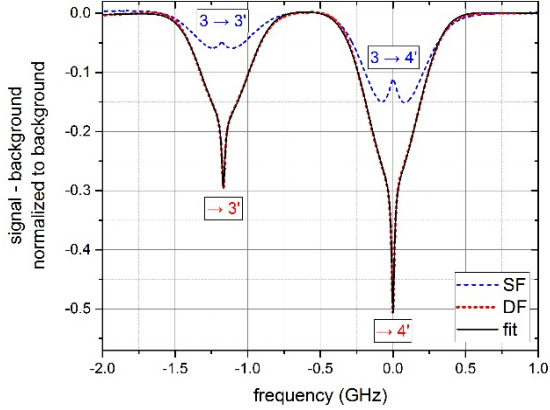

Fig. 10: Measurements of single-frequency (SF) and dual-frequency (DF) Doppler-free spectra through the microcell at 60°C. The laser power is about 200 μW after a single pass. The single-frequency spectrum shown here transitions obtained from the $F_1=3$ state. Here, spectra are fitted by a 2-Doppler plus 2-Lorentzian function with a quadratic background. Fit parameters for similar scans were used to deduce the height and the width of the Doppler-broadened and Doppler-free profiles. The frequency difference between the two Doppler-free peaks (1.16 GHz) was used to calibrate the frequency axis and then to estimate the linewidth of sub-Doppler resonances.

Fig.11 shows the results of the sub-Doppler resonance height measurements in the DF regime versus the reflection mirror position. The cell temperature is 42°C. The laser is connected to the $F_3=4$ excited state and the laser power is 45 μW. Measurements are compared to numerical calculations based on the density matrix equations from Appendix and Eq. (24). Spatial oscillations described in Section II-B are clearly visible.

The discrepancy between the experimental data and theory in Fig.11 can have several reasons. A first reason is the non-negligible optical thickness of the medium, neglected in our theory. A second reason is that the reflected beam undergoes intensity oscillations together with the oscillations of light absorption in the cell. This means that different positions of the mirror provide different combinations of forward and backward light beam intensities. In order to consider this effect correctly, a solution of the full system of Maxwell-Bloch equations would be needed. This approach is a quite complicated task and is not reasonable for our current study.

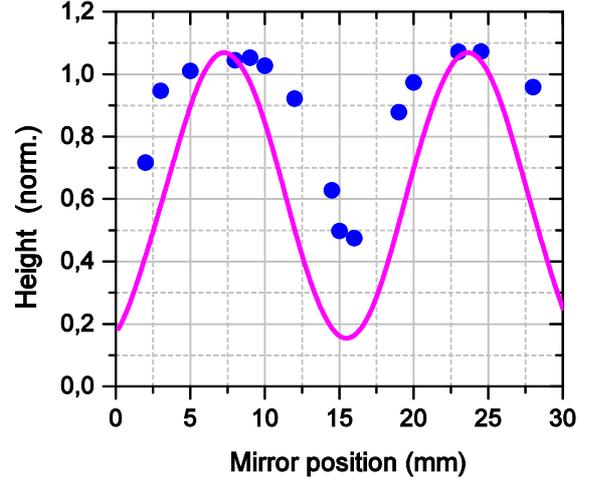

FIG. 11. Height of the sub-Doppler absorption spike normalized to the Doppler background height versus the distance between the mirror and the microcell. The solid pink line is the result of numerical calculations.

We would like to note again that these spatial oscillations of the sub-Doppler resonance height cannot be easily revealed in cm-scale cells, as explained in Section II-B and mentioned in [38]. Thus, short vapor cells are preferable for the observation of high-contrast enhanced-absorption spikes since the resonance contrast can be maximized by proper choice of the relative distance between the mirror and the vapor cell. Note that under the DF regime, the height of the sub-Doppler resonance can be even bigger than the height of the Doppler profile. Such a high relative contrast of the sub-Doppler resonance (>100%) is not possible in standard SF saturated absorption spectroscopy setups in which the relative contrast does not usually exceed 20-30%.

Figure 12 shows experimental results and calculations of the resonance linewidth and height in both SF and DF regimes versus the total laser power. In the DF regime, experimental results are reported for two different temperatures (42°C and 60°C). In experiments, intensities of the backward waves $E_{3,4}$ are not equal to those of incident waves $E_{1,2}$ (see sketch in Fig.1) due to absorption of light in the cell and different losses on the beam path. To take this into account, we consider the following relations in calculations: $I_1=I_2$, $I_3=I_4=0.5I_1$.

### C. Comparison with theory

*1. Linewidth measurements*

In experiments, the line-width of the resonance in the DF case is several times smaller than in the SF case and found to be closer to the natural linewidth when extrapolated at zero intensity. The resonance FWHM equals 59.1 MHz in the SF regime for a light power of about 70 μW, while it equals 16 MHz in the DF case for the same laser power (see Fig.12a). The narrowest sub-Doppler resonance in the DF regime is measured to be 13.75 MHz (P = 30.8 μW).

In the DF regime, experimental data are well fitted by numerical calculations. In the SF regime, the observed discrepancy between experience and theory (solid pink curve and red triangles in Fig.12a) can be explained by the same reasons than noted for Fig.11. In particular, the results for SF regime agree well under moderate light intensities ($I\sim I_{sat}$, i.e. $P_{light} \sim$ 1–10 μW) and become different with increase of the light intensity due to the influence of the high-order spatial harmonics of atom polarization. This influence is much more noticeable under the SF regime as predicted in the theory section. Additionally, experimental data for the resonance width might contain a residual Doppler broadening due to minor imperfections in the counter-propagating beams' alignment (e.g., see [55]).

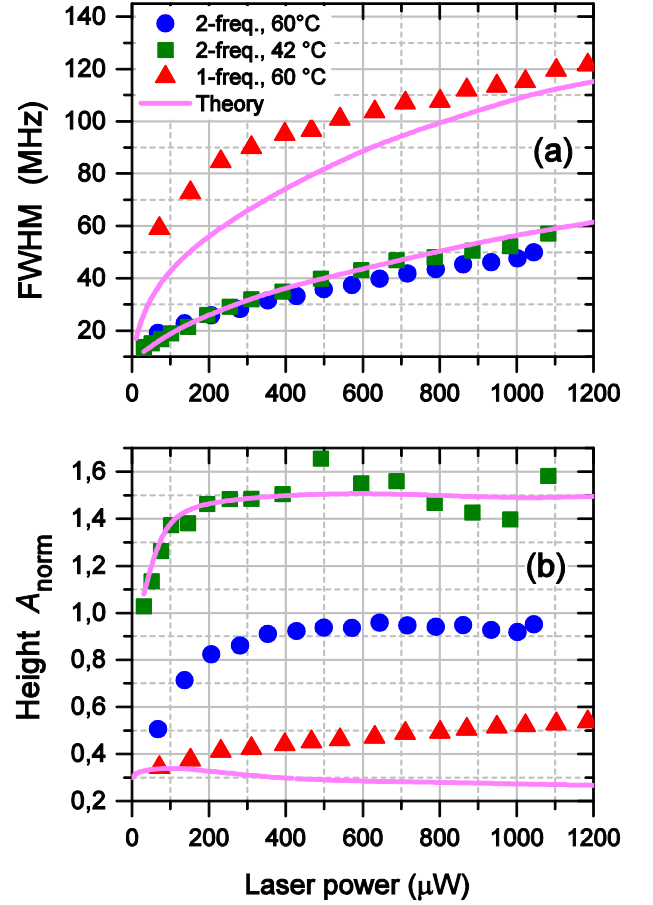

FIG. 12. (a) FWHM and (b) normalized height $A_{norm}$ of the sub-Doppler resonance as a function of the laser power under the SF and DF regimes of excitation. Comparison between experimental data and calculations (solid pink lines). The single-frequency resonance parameters are calculated for the transition $|1\rangle \rightarrow |3\rangle$. Other parameters are: $\alpha=\pi/2$, $\phi_{12}=0$, $B=0$.

In general, the resonance linewidth in both SF and DF regimes exhibits a well-known square-root-like dependence [1,2,55]:

$$\text{FWHM} \propto \gamma\sqrt{1+G}, \quad (25)$$

with $G$ being the effective saturation parameter. At the same time, the nonlinear resonance under the DF light-field configuration is significantly narrower. This can be explained by the influence of the openness of the optical transition in a simple two-level model. Indeed, in [56] the influence of the transition openness is governed by the parameter $\kappa$:

$$\kappa = \frac{\gamma_{eg}(2\Gamma + b\gamma)}{\Gamma(\Gamma + \gamma)}, \qquad (26)$$

with $\gamma_{eg}=\Gamma+\gamma/2$, and $\Gamma$ and $\gamma$ being the time-of-flight and spontaneous relaxation rates respectively. The coefficient of openness is determined as $b=1-\beta$, where $\beta$ is the branching ratio of the transition. It was shown in [56] that the transition openness significantly broadens the Bennett hole [1,2] and consequently the sub-Doppler resonance. Qualitatively, this effect can be understood by replacing $G$ in (25) with the product $\kappa G$. In the open-transition case, we have $\kappa \gg 1$ ($\Gamma \ll \gamma$, $\beta \sim 1$) and the nonlinear resonance experiences enhanced power broadening. The same effect is responsible for the larger linewidth of the saturated-absorption resonance under the SF regime, as shown in Fig. 11a. The DF regime with orthogonally polarized counter-propagating beams can be treated as a closed system of levels since no trap sublevels accumulate the atoms. Therefore, owing to condition $\kappa \approx 1$, the linewidth of the sub-Doppler resonance is significantly smaller in the DF regime.

*2. Height measurements*

Measurements show that the height of the resonance in the DF regime is about three times higher than in the SF case for a similar cell temperature of 60°C. In the DF case, experimental data are in good agreement with the theoretical model. The height of the resonance is maximized for a laser power of 600 μW at 60°C and can be 1.5 times higher than the wider Doppler background. This is totally impossible with the SF technique.

The influence of temperature on the resonance height is obvious. A higher temperature leads to increased optical thickness of the vapor and considerable light wave absorption in the cell. The increased absorption increases the imbalance between forward and backward light beams intensities. Note that this does not help to observe a higher-contrast normalized resonance (see Section II-C).

The dependency of the sub-Doppler resonance height on power in the DF regime (Fig.12b, green squares) can be qualitatively explained by a simplified two-level model, as reported in [56]. If the condition $\delta \gg$ FWHM is satisfied, the two dual-frequency beams act on different resonant groups of atoms and do not influence each other. As already discussed in Section II, atoms are in this condition pumped into dark states which do not interact with the laser. This situation is equivalent to an open transition and can be qualitatively treated as an open two-level atom model [56], with $\kappa \gg 1$. Close to the resonance center ($\delta \approx$ FWHM), atoms undergo excitation from both counter-propagating light beams. Due to the special experimental conditions (orthogonal linear polarizations, distance between the mirror and the cell properly-tuned), there are no trap states in atoms. Consequently, the real atomic system can be reduced to an effective closed two-level system with $\kappa \approx 1$.

Based on Eq. (19) from [56], we can come to the following expression for the normalized height of the sub-Doppler resonance:

$$A_{\text{norm}} = \frac{y_1 e^{y_2^2} \text{Erfc}[y_2]}{y_2 e^{y_1^2} \text{Erfc}[y_1]} - 1, \qquad (27)$$

with

$$y_1 = \frac{\gamma_{eg}}{\Delta_{\text{Dop}}}\sqrt{1+2\kappa G}, \quad y_2 = \frac{\gamma_{eg}}{\Delta_{\text{Dop}}}\sqrt{1+4G}, \qquad (28)$$

where $\Delta_{\text{Dop}}$ is the Doppler HWHM. Note that the first square root in (28) contains the coefficient $\kappa$ which stands for the open-system case, so that it should be considered to be much higher than unity. Erfc[$y$] in (27) is the complementary error function. The derived analytical expression (27) qualitatively agrees with the measurement results shown in Fig.12b (solid pink line for DF regime).

In the SF case, the height of the resonance is found to increase slightly up to 1.2 mW. For power values lower than 20 μW, both experimental and theoretical results are in agreement with a square-root-like growth. For higher power values, a discrepancy is observed. The theoretical curve demonstrates a very slow decrease whereas the experimental one continues its slow growth. This discrepancy may be explained by the influence of the high-order harmonics. As mentioned in the theory section, our experiments deal with quite high light-field intensities ($>> I_{sat} \sim 1$ mW). Therefore, high-order spatial harmonics of atom's polarization can influence the observed signals (see also [1,42]). Note that the high-order spatial harmonics have less noticeable influence in the DF regime due to the imbalance between forward and backward light beam intensities.

In general, the numerically calculated results for the DF regime are in good agreement with those obtained in the experiments. These results have demonstrated the significant quantitative advantage of the DF SDS technique over the commonly used SF technique. Moreover, the new technique has special interest for short vapor cells where spatial oscillations of the nonlinear resonance height can be well observed and involved to increase the resonance quality.

## IV. CONCLUSIONS

We have developed an extended and detailed theoretical model to describe the effect of high-contrast sub-Doppler resonances observed under the dual-frequency regime. This theory generalizes the previous qualitative simplified models proposed in [35,38]. The reported model considers the real structure of the atom and various complex physical phenomena. The theory helps to establish important requirements to optimize the sub-Doppler resonance contrast in experiments. These requirements are to use small length cells, to reduce the intensity imbalance between both counter-propagating beams, to operate at null magnetic field to save the Zeeman-CPT effect, to null the Raman detuning to allow the contribution of hyperfine CPT states, to use moderate laser power values of about 200-400 μW and to use orthogonal linearly polarized beams. The latter requirement is not strict since the use of elliptical polarizations could also give good results.

Experimental results were performed to validate the theoretical model using a Cs vapor micro-fabricated cell. Spatial oscillations of the sub-Doppler resonance height with translation of the reflection mirror position have been clearly demonstrated. This behavior is specific for a miniaturized cell since it results from the contribution of microwave hyperfine coherent population trapping states involved in the DF regime. The impact of the laser intensity and cell temperature on the sub-Doppler resonance properties has been studied in both SF and DF regimes. Experimental results are well explained by the theoretical model. Rigorous explanations were suggested to explain discrepancies.

Results demonstrate that the DF regime allows for the detection of narrower and higher contrast sub-Doppler resonances. This relevant feature could be of interest for the development of a microcell-based miniaturized optical frequency reference with high-stability performances. It could be interesting to study if this approach could be competitive with other miniature OFR architectures [57-60]. Note that by using Rb instead, the laser source could be based on a frequency-doubled telecom laser

[61,62], benefiting from the improved availability reliability and reproducibility of telecom-band lasers and optics.

## ACKNOWLEDGMENTS

This work has been supported by Région Bourgogne Franche-Comté and Labex FIRST-TF. We thank the platform Oscillator-Imp for the distribution of a reference hydrogen maser signal in the laboratory. The authors thank C. Rocher and P. Abbé (FEMTO-ST) for their help with experimental work and electronics. The work of D. Brazhnikov was supported by Russian Science Foundation (grant no. 17-72-20089).

## APPENDIX

Let us provide explicit expressions for the operators from (4). The free-atom Hamiltonian can be written as:

$$\hat{H}_0 = \sum_{F_a, m_a} \varepsilon_a |F_a, m_a\rangle \langle F_a, m_a|, \quad (A1)$$

where $\varepsilon_a$ are the energy of $a$-levels with $a = 1, 2, 3$ according to the notations introduced in Section II-A (see Fig.2).

The Hermitian light-atom interaction operator $\hat{V}_E$ has the form:

$$\hat{V}_E = \begin{pmatrix} \hat{0} & \hat{0} & \hat{V}_{31}^\dagger \\ \hat{0} & \hat{0} & \hat{V}_{32}^\dagger \\ \hat{V}_{31} & \hat{V}_{32} & \hat{0} \end{pmatrix}. \quad (A2)$$

In the rotating-wave and electric-dipole approximations, the matrix blocks in (A2) are:

$$\hat{V}_{31}(z,t) = -\hbar R_1 \hat{Y}_{31}^{(1)} e^{-i(\omega_1 t - k_1 z)}$$
$$-\hbar R_3 \hat{Y}_{31}^{(3)} e^{-i(\omega_1 t + k_1 z + \phi_1)}, \quad (A3)$$

$$\hat{V}_{32}(z,t) = -\hbar R_2 \hat{Y}_{32}^{(2)} e^{-i(\omega_2 t - k_2 z)}$$
$$-\hbar R_4 \hat{Y}_{32}^{(4)} e^{-i(\omega_2 t + k_2 z + \phi_2)}, \quad (A4)$$

with $R_i$ the Rabi frequencies and $\hat{Y}_{3a}^{(j)}$ ($a = 1, 2$, $j = 1$–$4$) the dimensionless interaction operators. According to the Wigner–Eckart theorem, we have:

$$\hat{Y}_{3a}^{(j)} = \boldsymbol{\xi}_j \cdot \hat{\mathbf{T}}^{3a}, \quad (A5)$$

where $\boldsymbol{\xi}_j$ is the $j$-wave polarization vector from (2), (3), and the $q$-components of vector operators $\hat{\mathbf{T}}$ are:

$$\hat{T}_q^{3a} = \sum_{m_3, m_a} (-1)^{F_3 - m_3} \begin{pmatrix} F_3 & 1 & F_a \\ -m_3 & q & m_a \end{pmatrix}$$
$$\times |F_3, m_3\rangle \langle F_a, m_a|, \quad (A6)$$

with $(\ldots)$ the 3jm-symbols [39].

The matrix form of the magnetic-field Hamiltonian $\hat{V}_B$ is:

$$\hat{V}_B = \hbar \Omega \begin{pmatrix} -\hat{F}_1 & \hat{0} & \hat{0} \\ \hat{0} & \hat{F}_2 & \hat{0} \\ \hat{0} & \hat{0} & \frac{g_3}{g_2}\hat{F}_3 \end{pmatrix}, \quad (A7)$$

where $\Omega$ is the Larmor frequency of the $F_2$ level.

The dimensionless operators $\hat{F}_a$ in (A7) stand for the $z$-projections of operators of total angular momentum in $F_a$ level. In the basis of eigenstates of the free-atom Hamiltonian $\hat{H}_0$, these operators have simple diagonal form:

$$\hat{F}_a = \sum_{m_a = -F_a, \ldots, F_a} m_a |F_a, m_a\rangle \langle F_a, m_a|, \quad (A8)$$

The part of operator $\hat{\mathfrak{R}}$ in (4) responsible for the spontaneous relaxation is:

$$\hat{\mathfrak{R}}^{\text{spon}} = \gamma(2F_3 + 1) \sum_{\substack{a=1,2 \\ q=0,\pm 1}} \beta_{3a} \hat{T}_q^{3a\dagger} \hat{\rho}^{aa} \hat{T}_q^{3a}, \quad (A9)$$

with $\gamma$ the spontaneous relaxation rate and $\beta_{3a}$ the branching ratios:

$$\beta_{3a} = (2J_e + 1)(2F_a + 1) \begin{Bmatrix} J_g & I_n & F_a \\ F_3 & 1 & J_e \end{Bmatrix}^2, \quad (A10)$$

Where $J_{e,g}$ are the total angular momenta of electrons in atom in excited (*e*) and ground (*g*) states, $I_n$ is the nuclear spin and $\{...\}$ stands for the 6j-symbol [39]. Obviously, $\beta_{31}+\beta_{32}=1$. For the $D_1$ line of Cs atom, we have $J_g=J_e=1/2$ and $I_n=7/2$.

The finite size of the light beams can be described by the time-of-flight relaxation term in $\hat{\Re}$:

$$\hat{\Re}^{\text{flight}} = \Gamma \left[ \hat{\rho}^{\text{isotr}} - \hat{\rho} \right], \quad (A11)$$

where $\Gamma$ is the time-of-flight relaxation rate, estimated as $\Gamma \approx \bar{\upsilon}/d$ with $\bar{\upsilon}$ the mean thermal velocity of atoms and $d$ the diameter of the light beams. As far as kinetic energy of atoms at temperature $T \approx 300$ K satisfy the condition $k_B T \gg \hbar \Delta_g$ (with $k_B$ the Boltzmann constant), the atom's magnetic sub-levels of both ground-state levels $|1\rangle$ and $|2\rangle$ (see Fig. 2) are populated equally and isotropically when atoms are beyond the light field. This initial atom's state is described by the matrix $\hat{\rho}^{\text{isotr}}$ in (A9):

$$\hat{\rho}^{\text{isotr}} = \frac{1}{2I_n+1} \sum_{a=1,2} |F_a, m_a\rangle\langle F_a, m_a|, \quad (A12)$$

where $(2I_n+1)=2(F_1+F_2+1)$ is the total number of magnetic sub-levels in the ground state of atom.

Following the harmonic expansions (6)-(8), we can write for the optical coherences (see also [5,38]):

$$\hat{\rho}_{13}(z,t) = e^{i\omega_1 t}\left( \hat{\rho}_{13}^{(-1)} e^{-ik_1 z} + \hat{\rho}_{13}^{(+1)} e^{i(k_1 z+\phi_1)} \right.$$
$$\left. + \hat{\rho}_{13}^{(-21)} e^{-i(2k_2-k_1)z} + \hat{\rho}_{13}^{(+21)} e^{i(2k_2-k_1)z+i\phi_1} \right), (A13)$$

$$\hat{\rho}_{23}(z,t) = e^{i\omega_2 t}\left( \hat{\rho}_{23}^{(-2)} e^{-ik_2 z} + \hat{\rho}_{23}^{(+2)} e^{i(k_2 z+\phi_2)} \right.$$
$$\left. + \hat{\rho}_{23}^{(-12)} e^{-i(2k_1-k_2)z} + \hat{\rho}_{23}^{(+12)} e^{i(2k_1-k_2)z+i\phi_2} \right). (A14)$$

Similar expansions can be written for Hermitian conjugate matrices $\hat{\rho}_{31} = \hat{\rho}_{13}^\dagger$ and $\hat{\rho}_{32} = \hat{\rho}_{23}^\dagger$.

The static magnetic field *B*, if present in the vapor cell, is assumed to be small enough to satisfy the condition $\Omega \ll \gamma$. This allows to take into account the magnetic field influence only on the ground-state levels. This means that the Zeeman splitting of the saturated-absorption resonance is not considered, while the influence of the *B*-field on creation of the Zeeman-CPT effect is considered.

All the listed assumptions help us to exclude the optical coherences (A13), (A14) and conjugate terms from the final system of equations to be solved numerically (optical Bloch equations). In particular, we have for the matrix harmonics of the ground state $F_1$:

$$\left( \Gamma + R_1^2 L_1^{(-)*} \hat{\Upsilon}_{13}^{(1)} \hat{\Upsilon}_{31}^{(1)} + R_3^2 L_1^{(+)*} \hat{\Upsilon}_{13}^{(3)} \hat{\Upsilon}_{31}^{(3)} \right) \hat{\rho}_{11}^{(0)}$$
$$+ \hat{\rho}_{11}^{(0)} \left( R_1^2 L_1^{(-)} \hat{\Upsilon}_{13}^{(1)} \hat{\Upsilon}_{31}^{(1)} + R_3^2 L_1^{(+)} \hat{\Upsilon}_{13}^{(3)} \hat{\Upsilon}_{31}^{(3)} \right)$$
$$- 2\gamma_{eg} S_{11}^{(-)} \hat{\Upsilon}_{13}^{(1)} \hat{\rho}_{33}^{(0)} \hat{\Upsilon}_{31}^{(1)} - 2\gamma_{eg} S_{31}^{(+)} \hat{\Upsilon}_{13}^{(3)} \hat{\rho}_{33}^{(0)} \hat{\Upsilon}_{31}^{(3)}$$
$$+ R_1 R_2 L_1^{(-)*} \hat{\Upsilon}_{13}^{(1)} \hat{\Upsilon}_{32}^{(2)} \hat{\rho}_{21}^{(+)} + R_3 R_4 L_1^{(+)} \hat{\rho}_{12}^{(+)} \hat{\Upsilon}_{23}^{(4)} \hat{\Upsilon}_{31}^{(3)} e^{-i\phi_{12}}$$
$$+ R_1 R_2 L_1^{(-)} \hat{\rho}_{12}^{(-)} \hat{\Upsilon}_{23}^{(2)} \hat{\Upsilon}_{31}^{(1)} + R_3 R_4 L_1^{(+)*} \hat{\Upsilon}_{13}^{(3)} \hat{\Upsilon}_{32}^{(4)} \hat{\rho}_{21}^{(-)} e^{i\phi_{12}}$$
$$- i\Omega \left[ \hat{F}_1, \hat{\rho}_{11}^{(0)} \right] - \hat{\Re}_{11}^{\text{spon}} \{ \hat{\rho}_{33}^{(0)} \} = \Gamma \hat{\rho}_{11}^{\text{isotr}}, \quad (A15)$$

$$\left( \Gamma + 2ik_{12}\upsilon + R_3^2 M_1^{(+)*} \hat{\Upsilon}_{13}^{(3)} \hat{\Upsilon}_{31}^{(3)} \right) \hat{\rho}_{11}^{(+)}$$
$$+ R_1^2 M_1^{(-)} \hat{\rho}_{11}^{(+)} \hat{\Upsilon}_{13}^{(1)} \hat{\Upsilon}_{31}^{(1)} - i\Omega \left[ \hat{F}_1, \hat{\rho}_{11}^{(+)} \right] - \hat{\Re}_{11}^{\text{spon}} \{ \hat{\rho}_{33}^{(+)} \}$$
$$- R_1^2 M_1^{(-)} \hat{\Upsilon}_{13}^{(1)} \hat{\rho}_{33}^{(+)} \hat{\Upsilon}_{31}^{(1)} - R_3^2 M_1^{(+)*} \hat{\Upsilon}_{13}^{(3)} \hat{\rho}_{33}^{(+)} \hat{\Upsilon}_{31}^{(3)}$$
$$+ R_1 R_2 M_1^{(-)} \hat{\rho}_{12}^{(+)} \hat{\Upsilon}_{23}^{(2)} \hat{\Upsilon}_{31}^{(1)}$$
$$+ R_3 R_4 M_1^{(+)*} \hat{\Upsilon}_{13}^{(3)} \hat{\Upsilon}_{32}^{(4)} \hat{\rho}_{21}^{(+)} e^{i\phi_{12}} = 0. \quad (A16)$$

Here and after square brackets with comma $[...,...]$ stand for the commutation operation of two matrices. Also, as long as $\hat{\rho}_{11}^{(-)} = \hat{\rho}_{11}^{(+)\dagger}$, the equation for matrix $\hat{\rho}_{11}^{(-)}$ can be easily derived from (A16) just by means of Hermitian conjugation of all terms. Therefore, we do not show it here.

Similarly, we get for the ground state $F_2$:

$$\left(\Gamma + R_2^2 L_2^{(-)*}\hat{Y}_{23}^{(2)}\hat{Y}_{32}^{(2)} + R_4^2 L_2^{(+)*}\hat{Y}_{23}^{(4)}\hat{Y}_{32}^{(4)}\right)\hat{\rho}_{22}^{(0)}$$
$$+\hat{\rho}_{22}^{(0)}\left(R_2^2 L_2^{(-)}\hat{Y}_{23}^{(2)}\hat{Y}_{32}^{(2)} + R_4^2 L_2^{(+)}\hat{Y}_{23}^{(4)}\hat{Y}_{32}^{(4)}\right)$$
$$-2\gamma_{eg}S_{22}^{(-)}\hat{Y}_{23}^{(2)}\hat{\rho}_{33}^{(0)}\hat{Y}_{32}^{(2)} - 2\gamma_{eg}S_{42}^{(+)}\hat{Y}_{23}^{(4)}\hat{\rho}_{33}^{(0)}\hat{Y}_{32}^{(4)}$$
$$+R_1 R_2 L_2^{(-)*}\hat{Y}_{23}^{(2)}\hat{Y}_{31}^{(1)}\hat{\rho}_{12}^{(-)} + R_3 R_4 L_2^{(+)}\hat{\rho}_{21}^{(-)}\hat{Y}_{13}^{(3)}\hat{Y}_{32}^{(4)}e^{i\phi_{12}}$$
$$+R_1 R_2 L_2^{(-)}\hat{\rho}_{12}^{(+)}\hat{Y}_{13}^{(1)}\hat{Y}_{32}^{(2)} + R_3 R_4 L_2^{(+)*}\hat{Y}_{23}^{(4)}\hat{Y}_{31}^{(3)}\hat{\rho}_{12}^{(+)}e^{-i\phi_{12}}$$
$$+i\Omega\left[\hat{F}_2, \hat{\rho}_{22}^{(0)}\right] - \hat{\mathfrak{R}}_{22}^{\text{spon}}\left\{\hat{\rho}_{33}^{(0)}\right\} = \Gamma\,\hat{\rho}_{22}^{\text{isotr}}, \quad \text{(A17)}$$

$$\left(\Gamma + 2ik_{12}\upsilon + R_2^2 M_2^{(-)*}\hat{Y}_{23}^{(2)}\hat{Y}_{32}^{(2)}\right)\hat{\rho}_{22}^{(+)}$$
$$+R_4^2 M_2^{(+)}\hat{\rho}_{22}^{(+)}\hat{Y}_{23}^{(4)}\hat{Y}_{32}^{(4)} + i\Omega\left[\hat{F}_2, \hat{\rho}_{22}^{(+)}\right] - \hat{\mathfrak{R}}_{22}^{\text{spon}}\left\{\hat{\rho}_{33}^{(+)}\right\}$$
$$-R_2^2 M_2^{(-)*}\hat{Y}_{23}^{(2)}\hat{\rho}_{33}^{(+)}\hat{Y}_{32}^{(2)} - R_4^2 M_2^{(+)}\hat{Y}_{23}^{(4)}\hat{\rho}_{33}^{(+)}\hat{Y}_{32}^{(4)}$$
$$+R_1 R_2 M_2^{(-)*}\hat{Y}_{23}^{(2)}\hat{Y}_{31}^{(1)}\hat{\rho}_{12}^{(+)}$$
$$+R_3 R_4 M_2^{(+)}\hat{\rho}_{21}^{(+)}\hat{Y}_{13}^{(3)}\hat{Y}_{32}^{(4)}e^{i\phi_{12}} = 0. \quad \text{(A18)}$$

The Hermitian conjugation of (A18) leads to the equation for $\hat{\rho}_{22}^{(-)}$.

Following equations are for the upper state:

$$\left(\Gamma + \gamma + R_1^2 L_1^{(-)}\hat{Y}_{31}^{(1)}\hat{Y}_{13}^{(1)} + R_2^2 L_2^{(-)}\hat{Y}_{32}^{(2)}\hat{Y}_{23}^{(2)}\right.$$
$$\left.+R_3^2 L_1^{(+)}\hat{Y}_{31}^{(3)}\hat{Y}_{13}^{(3)} + R_4^2 L_2^{(+)}\hat{Y}_{32}^{(4)}\hat{Y}_{23}^{(4)}\right)\hat{\rho}_{33}^{(0)}$$
$$+\hat{\rho}_{33}^{(0)}\left(R_1^2 L_1^{(-)*}\hat{Y}_{31}^{(1)}\hat{Y}_{13}^{(1)} + R_2^2 L_2^{(-)*}\hat{Y}_{32}^{(2)}\hat{Y}_{23}^{(2)}\right.$$
$$\left.+R_3^2 L_1^{(+)*}\hat{Y}_{31}^{(3)}\hat{Y}_{13}^{(3)} + R_4^2 L_2^{(+)*}\hat{Y}_{32}^{(4)}\hat{Y}_{23}^{(4)}\right)$$
$$-2\gamma_{eg}S_{11}^{(-)}\hat{Y}_{31}^{(1)}\hat{\rho}_{11}^{(0)}\hat{Y}_{13}^{(1)} - 2\gamma_{eg}S_{31}^{(+)}\hat{Y}_{31}^{(3)}\hat{\rho}_{11}^{(0)}\hat{Y}_{13}^{(3)}$$
$$-2\gamma_{eg}S_{22}^{(-)}\hat{Y}_{32}^{(2)}\hat{\rho}_{22}^{(0)}\hat{Y}_{23}^{(2)} - 2\gamma_{eg}S_{42}^{(+)}\hat{Y}_{32}^{(4)}\hat{\rho}_{22}^{(0)}\hat{Y}_{23}^{(4)}$$
$$-R_1 R_2 \left(L_1^{(-)} + L_2^{(-)*}\right)\hat{Y}_{31}^{(1)}\hat{\rho}_{12}^{(-)}\hat{Y}_{23}^{(2)}$$
$$-R_1 R_2 \left(L_2^{(-)} + L_1^{(-)*}\right)\hat{Y}_{32}^{(2)}\hat{\rho}_{21}^{(+)}\hat{Y}_{13}^{(1)}$$
$$-R_3 R_4 \left(L_2^{(+)} + L_1^{(+)*}\right)\hat{Y}_{32}^{(4)}\hat{\rho}_{21}^{(-)}\hat{Y}_{13}^{(3)}e^{i\phi_{12}}$$
$$-R_3 R_4 \left(L_1^{(+)} + L_2^{(+)*}\right)\hat{Y}_{31}^{(3)}\hat{\rho}_{12}^{(+)}\hat{Y}_{23}^{(4)}e^{-i\phi_{12}}$$
$$+i(g_e/g_2)\Omega\left[\hat{F}_3, \hat{\rho}_{33}^{(0)}\right] = 0. \quad \text{(A19)}$$

$$\left(\Gamma + \gamma + 2ik_{12}\upsilon + R_1^2 M_1^{(-)}\hat{Y}_{31}^{(1)}\hat{Y}_{13}^{(1)}\right.$$
$$\left.+R_4^2 M_2^{(+)}\hat{Y}_{32}^{(4)}\hat{Y}_{23}^{(4)}\right)\hat{\rho}_{33}^{(+)} + i(g_e/g_2)\Omega\left[\hat{F}_3, \hat{\rho}_{33}^{(+)}\right]$$
$$+\hat{\rho}_{33}^{(+)}\left(R_2^2 M_2^{(-)*}\hat{Y}_{32}^{(2)}\hat{Y}_{23}^{(2)} + R_3^2 M_1^{(+)*}\hat{Y}_{31}^{(3)}\hat{Y}_{13}^{(3)}\right)$$
$$-R_1^2 M_1^{(-)}\hat{Y}_{31}^{(1)}\hat{\rho}_{11}^{(+)}\hat{Y}_{13}^{(1)} - R_3^2 M_1^{(+)*}\hat{Y}_{31}^{(3)}\hat{\rho}_{11}^{(+)}\hat{Y}_{13}^{(3)}$$
$$-R_2^2 M_2^{(-)*}\hat{Y}_{32}^{(2)}\hat{\rho}_{22}^{(+)}\hat{Y}_{23}^{(2)} - R_4^2 M_2^{(+)}\hat{Y}_{32}^{(4)}\hat{\rho}_{22}^{(+)}\hat{Y}_{23}^{(4)}$$
$$-R_1 R_2 \left(M_1^{(-)} + M_2^{(-)*}\right)\hat{Y}_{31}^{(1)}\hat{\rho}_{12}^{(+)}\hat{Y}_{23}^{(2)}$$
$$-R_3 R_4 \left(M_1^{(+)*} + M_2^{(+)}\right)\hat{Y}_{32}^{(4)}\hat{\rho}_{21}^{(+)}\hat{Y}_{13}^{(3)}e^{i\phi_{12}} = 0. \quad \text{(A20)}$$

The Hermitian conjugated equation (A20) gives the equation for $\hat{\rho}_{33}^{(-)}$.

Finally, for the low frequency coherences, we get:

$$\left(\Gamma + i[\delta_R + k_{12}\upsilon] + R_1^2 M_2^{(-)*}\hat{Y}_{13}^{(1)}\hat{Y}_{31}^{(1)}\right.$$
$$\left.+R_3^2 L_2^{(+)*}\hat{Y}_{13}^{(3)}\hat{Y}_{31}^{(3)}\right)\hat{\rho}_{12}^{(+)} - i\Omega\left(\hat{F}_1\hat{\rho}_{12}^{(+)} + \hat{\rho}_{12}^{(+)}\hat{F}_2\right)$$
$$+\hat{\rho}_{12}^{(+)}\left(R_2^2 M_1^{(-)}\hat{Y}_{23}^{(2)}\hat{Y}_{32}^{(2)} + R_4^2 L_1^{(+)}\hat{Y}_{23}^{(4)}\hat{Y}_{32}^{(4)}\right)$$
$$+R_1 R_2 M_1^{(-)}\hat{\rho}_{11}^{(+)}\hat{Y}_{13}^{(1)}\hat{Y}_{32}^{(2)} + R_3 R_4 L_1^{(+)}\hat{\rho}_{11}^{(0)}\hat{Y}_{13}^{(3)}\hat{Y}_{32}^{(4)}e^{i\phi_{12}}$$
$$+R_1 R_2 M_2^{(-)*}\hat{Y}_{13}^{(1)}\hat{Y}_{32}^{(2)}\hat{\rho}_{22}^{(+)} + R_3 R_4 L_2^{(+)*}\hat{Y}_{13}^{(3)}\hat{Y}_{32}^{(4)}\hat{\rho}_{22}^{(0)}e^{i\phi_{12}}$$
$$-R_1 R_2 \left(M_1^{(-)} + M_2^{(-)*}\right)\hat{Y}_{13}^{(1)}\hat{\rho}_{33}^{(+)}\hat{Y}_{32}^{(2)}$$
$$-R_3 R_4 \left(L_1^{(+)} + L_2^{(+)*}\right)\hat{Y}_{13}^{(3)}\hat{\rho}_{33}^{(0)}\hat{Y}_{32}^{(4)}e^{i\phi_{12}} = 0. \quad \text{(A21)}$$

$$\left(\Gamma + i[\delta_R - k_{12}\upsilon] + R_1^2 L_2^{(-)*}\hat{Y}_{13}^{(1)}\hat{Y}_{31}^{(1)}\right.$$
$$\left.+R_3^2 M_2^{(+)*}\hat{Y}_{13}^{(3)}\hat{Y}_{31}^{(3)}\right)\hat{\rho}_{12}^{(-)} - i\Omega\left(\hat{F}_1\hat{\rho}_{12}^{(-)} + \hat{\rho}_{12}^{(-)}\hat{F}_2\right)$$
$$+\hat{\rho}_{12}^{(-)}\left(R_2^2 L_1^{(-)}\hat{Y}_{23}^{(2)}\hat{Y}_{32}^{(2)} + R_4^2 M_1^{(+)}\hat{Y}_{23}^{(4)}\hat{Y}_{32}^{(4)}\right)$$
$$+R_1 R_2 L_1^{(-)}\hat{\rho}_{11}^{(0)}\hat{Y}_{13}^{(1)}\hat{Y}_{32}^{(2)} + R_3 R_4 M_1^{(+)}\hat{\rho}_{11}^{(-)}\hat{Y}_{13}^{(3)}\hat{Y}_{32}^{(4)}e^{i\phi_{12}}$$
$$+R_1 R_2 L_2^{(-)*}\hat{Y}_{13}^{(1)}\hat{Y}_{32}^{(2)}\hat{\rho}_{22}^{(0)} + R_3 R_4 M_2^{(+)*}\hat{Y}_{13}^{(3)}\hat{Y}_{32}^{(4)}\hat{\rho}_{22}^{(-)}e^{i\phi_{12}}$$
$$-R_1 R_2 \left(L_1^{(-)} + L_2^{(-)*}\right)\hat{Y}_{13}^{(1)}\hat{\rho}_{33}^{(0)}\hat{Y}_{32}^{(2)}$$
$$-R_3 R_4 \left(M_1^{(+)} + M_2^{(+)*}\right)\hat{Y}_{13}^{(3)}\hat{\rho}_{33}^{(-)}\hat{Y}_{32}^{(4)}e^{i\phi_{12}} = 0. \quad \text{(A22)}$$

Since $\hat{\rho}_{21}^{(-)} = \hat{\rho}_{12}^{(+)\dagger}$ and $\hat{\rho}_{21}^{(+)} = \hat{\rho}_{12}^{(-)\dagger}$, the other two equations can be obtained directly by Hermitian conjugation of the last two equations.

In (A15)-(A22), several new notations have been introduced. In particular, the saturation parameters are:

$$S_{n1}^{(\pm)} = \frac{R_n^2}{\gamma_{eg}^2 + \left(\delta + \frac{\delta_R}{2} \pm k_1 \upsilon\right)^2} \quad (n = 1, 3), \quad (A23)$$

$$S_{n2}^{(\pm)} = \frac{R_n^2}{\gamma_{eg}^2 + \left(\delta - \frac{\delta_R}{2} \pm k_2 \upsilon\right)^2} \quad (n = 2, 4), \quad (A24)$$

and the complex Lorentzians are:

$$L_1^{(\pm)} = \left[\gamma_{eg} + i\left(\delta + \frac{\delta_R}{2} \pm k_1 \upsilon\right)\right]^{-1}, \quad (A25)$$

$$L_2^{(\pm)} = \left[\gamma_{eg} + i\left(\delta - \frac{\delta_R}{2} \pm k_2 \upsilon\right)\right]^{-1}, \quad (A26)$$

$$M_1^{(\pm)} = \left[\gamma_{eg} + i\left(\delta + \frac{\delta_R}{2} \pm (2k_2 - k_1)\upsilon\right)\right]^{-1}, \quad (A27)$$

$$M_2^{(\pm)} = \left[\gamma_{eg} + i\left(\delta - \frac{\delta_R}{2} \pm (2k_1 - k_2)\upsilon\right)\right]^{-1}. \quad (A28)$$

The total excited state population of the atom as well as the absorption coefficient (13) can be obtained by numerically solving the equations (A15)-(A22).